# Evidence for conventional superconductivity in $Sr_{0.1}Bi_2Se_3$ from high pressure studies

K. Manikandan,[1+] Shruti,[2+] P. Neha,[2] G. Kalai Selvan,[1] B. Wang,[3] Y. Uwatoko,[3] K. Ishigaki,[3] R. Jha,[4] V.P.S. Awana,[4] S.Arumugam,[1*] S. Patnaik[2*]

[1] Centre for High Pressure Research, School of Physics, Bharathidasan University,

Tiruchirappalli - 620024, India

[2] School of Physical Sciences, Jawaharlal Nehru University, New Delhi, India-110067

[3] Institute for Solid State Physics, University of Tokyo, Kashiwa, Chiba 277-8581, Japan

[4] Quantum Phenomena and Applications Division, National Physical Laboratory (CSIR)

Dr. K. S. Krishnan Road, New Delhi-110012, India

## Abstract

$Sr_xBi_2Se_3$ is recently reported to be a superconductor derived from topological insulator $Bi_2Se_3$. It shows a maximum resistive $T_c$ of 3.25 K at ambient pressure. We report magnetic (upto 1 GPa) and transport properties (upro 8 Gpa) under pressure for single crystalline $Sr_{0.1}Bi_2Se_3$ superconductor. Magnetic measurements show that $T_c$ decreases from ~2.6 K (0 GPa) to ~1.9 K (0.81 GPa). Similar behavior is observed in transport properties as well without much change in the metallic characteristics in normal state resistivity. No reentrant superconducting phase (Physical Review B **93**, 144514 (2016)) is observed at high pressure. Normal state resistivity near $T_c$ is explained by Fermi liquid model. Above 100 K, a polaronic hopping conduction mechanism with two parallel channels for current flow is indicated. Band structure calculations indicate decreasing density of states at Fermi level with pressure. In consonance with transition temperature suppression in conventional BCS low $T_c$ superconductors, the pressure effect in $Sr_xBi_2Se_3$ is well accounted by pressure induced band broadening.

[*]Corresponding authors: spatnaik@mail.jnu.ac.in, sarumugam1963@yahoo.com

[+] First and second authors have made equal contribution to this paper

# 1. Introduction

Topological insulators are novel quantum states of matter that have attracted significant attention due to their peculiar property of conducting surface or edge states emerging out of non-trivial topology of electronic band structure [1-3]. $Bi_2Te_3$, $Bi_2Se_3$ and $Sb_2Te_3$ belong to the group of 3D topological insulators with associated single Dirac cone surface states [4 - 7]. In particular, superconductors derived from topological insulators (TI) have been predicted to host Majorana fermions (particles that are their own anti-particles) which can be of great significance towards realizing quantum computation [8, 9]. Through judicious doping unto TIs, low $T_c$ superconductivity has been observed in $Cu_xBi_2Se_3$, $Pd_xBi_2Te_3$ and $Sn_{1-x}In_xTe$ and theories based on unconventional superconductivity have been proposed [10 - 14]. A key feature of topological superconductivity is the experimental proof of p-wave superconductivity that has eluded thus far. Apart from chemical doping, an alternative approach to induce superconductivity is to tune the electronic structure by applying high pressure. The electrical and structural properties of $Bi_2Te_3$ and $Sb_2Te_3$ as well as their electronic band structure, have been studied at high pressures in great detail, and several recent studies have provided evidence for pressure induced superconductivity in these systems [15 - 17]. On the other hand, both in case of $Cu_xBi_2Se_3$, $Sn_{1-x}In_xTe$, it is confirmed that $T_c$ decreases as a function of pressure and, it gets completely suppressed at ~ 6 GPa in $Cu_xBi_2Se_3$ [18]. Moreover, pressure induced superconductivity has been observed in undoped $Bi_2Se_3$ above 11 GPa that achieves a maximum $T_c$ of 7 K above 30 GPa and this value is maintained upto 50 GPa pressure [19]. Very recently superconductivity is reported in Sr intercalated topological insulator $Bi_2Se_3$ with $T_c$ ~ 2.9 K at ambient pressure and large superconducting shielding fraction [20, 21]. Topological surface states well separated from bulk bands have been confirmed in this compound and it is thought to be another promising candidate for topological superconductivity [22]. Several recent high pressure studies have reported evidence for unconventional superconductivity from intricate analysis of basal plane upper critical field as a function of pressure [23 - 25]. However, the key question of experimental evidence for p-wave superconductivity has remained unanswered.

In this paper, we revisit the effect of hydrostatic pressure on the superconducting properties of Sr intercalated $Bi_2Se_3$ single crystals. The analysis incorporates pressure induced changes both from magnetization and resistivity data. We find systematic $T_c$ suppression on the application of pressure upto 1.85 GPa. Further, temperature dependence of normal state resistivity using cubic press is measured and the normal state resistivity behavior is explained. No reentrant superconducting phase (upto 8 GPa) is observed. Fermi liquid behaviour is seen to evolve in low temperature range. The electronic states evolving out of pressure effect in $Sr_{0.1}Bi_2Se_3$ have also been analyzed using VASP 5.4 code and our data are in accordance with conventional BCS pairing mechanism for superconductivity in Sr intercalated $Bi_2Se_3$.

2. Experiment

The preparation method of single crystal of $Sr_{0.1}Bi_2Se_3$ is described elsewhere [21]. The compound crystallizes in rhombohedral structure with space group $R-3m$. The structural details are given in reference 21. Magnetization measurements at various pressures were performed using Magnetic Property Measurement System (MPMS, Quantum Design, USA). The external pressure was generated upto 1 GPa by a clamp type miniature hydrostatic pressure cell which is made of nonmagnetic Cu–Be alloy. The fluorinert FC#70 and FC#77 (1:1) mixture was used as a pressure transmitting medium and the in-situ pressure (P) was estimated from the superconducting transition of pure Sn as a manometer. Temperature dependence of magnetization M(T) was recorded upon zero field cooling at various pressures under external magnetic field of 20 Oe . The resistivity measurements upto 2GPa was done with HPC-33 Piston type pressure cell in Physical Property Measurements System (PPMS-14T, Quantum Design). Hydrostatic pressures was applied using BeCu/NiCrAl clamped piston- cylinder cell immersed in Fluorinert (FC70:FC77=1:1) pressure transmitting medium. A cubic anvil device was used for electrical resistivity measurements for various pressures from 2 to 8 GPa. The sample was immersed in a pressure medium of Daphne #7373 oil to maintain the hydrostatic pressure and encapsulated in a Teflon cell, surrounded by a pyrophyllite block. This block was evenly compressed from six directions using six tungsten carbide (WC) anvils. The six WC anvils crush the pyrophyllite gasket and compress the Teflon cell from six directions equally and the hydrostatic nature of the pressure is maintained beyond the solidification of the pressure medium. Furthermore, the pressure is controlled and kept constant during warming and cooling cycle of the cubic press device. The pressure was calibrated using resistive transitions of Bi I–II (2.55 GPa), Bi II–III (2.7 GPa) and III–IV (7.7 GPa) at room temperature [26].

## 3. Results and Discussion

Figure 1 shows the temperature dependent resistivity of $Sr_{0.1}Bi_2Se_3$ measured using both clamp type piston cylinder pressure cell (upto 2 GPa) and cubic press experimental setup (from 2-8 GPa). The low temperature region of R(T) under various pressures is shown as an inset of figure 1. It is found that the ambient pressure the sample shows a $T_c^{onset}$ of 3.25 K. With the application of pressure, superconducting transition shifts to low temperature in transport measurements. However, $T_c^{onset}$ is not observed above 0.35 GPa due to the temperature limitation of 2 K in the cubic press experimental setup for R(T) measurements. Figure 2 shows temperature dependent DC magnetization measurements in zero field cool (ZFC) mode under various hydrostatic pressures such as 0, 0.18, 0.51, 0.81, 0.99 GPa on $Sr_{0.1}Bi_2Se_3$ sample. The main panel of figure 2 shows the magnetization (M(T)) subtracted with the magnetization at 3 K (M(3 K)) to compare the superconducting transition temperature ($T_c$) at various pressure values. At ambient pressure $Sr_{0.1}Bi_2Se_3$ shows a magnetization $T_c$ of 2.67 K which is lower by ~ 0.6K compared to that measured from resistance measurements. The magnetization data suggests that bulk superconductivity with very large superconducting volume fraction can be achieved by Sr intercalation in topological insulator $Bi_2Se_3$. The inset of figure 2 shows the expanded view that shows the pressure dependence of Sn $T_c$ (marked inside the ellipsoidal curve). Thus $T_c$ of $Sr_{0.1}Bi_2Se_3$ and Sn are superimposed at less than 3 K under pressure upto ~ 1 GPa. With application of external pressure, initially $T_c$ varies marginally upto 0.18GPa and then shifts to lower temperature. Evidently, $T_c^{onset}$ observed from R(T, P) measurements is higher than the $T_c$ observed in M(T, P) measurements.

Figure 3 shows the pressure dependence of $T_c$ obtained from M(T) measurements and it is extrapolated using a polynomial fit to indicate the possible suppression of $T_c$ at critical pressure of 1.85 GPa. It is found from the figure 3 that the $T_c$ decreases ($dT_c/dP$ = -1.14 K/GPa) and gets suppressed at 1.85 GPa. Similar suppression was also seen with resistivity data but the $dT_c/dP$ was found to be = -1.04 K/GPa. Clearly, $T_c$ is found to decrease and get suppressed for both $Cu_xBi_2Se_3$ [18] and $Sr_{0.1}Bi_2Se_3$ compounds under pressure. Also, pressure dependence of $T_c$ is ~3 times higher and the critical pressure is ~3 times less in $Sr_{0.1}Bi_2Se_3$ compared to $Cu_xBi_2Se_3$ sample [18]. A recent paper on high pressure studies in $Sr_{0.05}Bi_2Se_3$ [23] suggests a structural phase transition from rhombohedral to monoclinic (above 6 GPa) and to tetragonal (above 25 GPa). It reports suppression of superconductivity with pressure along with a re-entrant superconducting phase above 6 GPa. We do not find evidence for the high pressure superconducting phase. In fact it is possible, that the re-entrant phase seen earlier would probably relate to structural phase transition of $Bi_2Se_3$ rather than due to Sr intercalation effects. To elucidate normal state properties, inset of figure 3 shows the pressure dependent resistivity at 3.5 K and 300 K upto the maximum pressure of 8 GPa. It reveals that the normal state resistivity shows increase upto 2 GPa and followed by a rather steep increase upto 8 GPa.

Towards a qualitative understanding of our data under BCS model, $T_c$ can be written as $T_c \sim \Theta_D \exp[-1 / N(E_F)\nu_{eff}]$ where $\Theta_D$ is the Debye temperature, $N(E_F)$ is the electronic density of states at the Fermi energy and $\nu_{eff}$ is the effective electron –phonon interaction parameter [18, 27]. Since $\Theta_D$ commonly increases with pressure due to phonon hardening, the observed decrease in $T_c$ could be related to the decrease in $N(E_F)$ as reported in $Cu_xBi_2Se_3$ compound [18]. Indeed, the effective interaction parameter can also be pressure dependent but in conventional low $T_c$ superconductors, this decreasing $T_c$ with pressure is easily explained by pressure induced band broadening leading to changes in $N(E_F)$ [27]. We note from figure 1 and inset in figure 3 that normal state resistivity increases with increasing pressure. This is also indicative of decreasing $N((E_F)$. In figure 4 we summarize the electronic band structure study of pure $Bi_2Se_3$, $Sr_{0.1}Bi_2Se_3$ at ambient pressure and $Sr_{0.2}Bi_2se_3$ at 2 GPa pressure. The impact of intercalated 'Sr' atom in topological insulator $Bi_2Se_3$ and the effect of external pressure can be deciphered from the density of states at the Fermi level. The band structure and density of states are calculated by using density functional theory (DFT) by employing Vienna Ab-initio Simulation Package (VASP 5.4). The DFT calculations were performed using GGA-PBE (Generalized Gradient

Approximation-Perdew Burke Ernzerhof) approximations to define exchange correlation energy. Typically 18×18×18 mesh points were used with K-point grid of 0.098×0.098×0.098 to achieve optimal convergence. For the calculation of band structure and density of state, we employed PAW (projected augmented waves and plane wave basis set of 315 eV cut off. The band structure calculated in highly symmetric direction clearly shows crossing of bands about Fermi energy about Gamma point(G) for 'Sr' intercalated $Bi_2Se_3$ while for parent compound $Bi_2Se_3$ the bands do not cross the Fermi energy at Gamma point as expected for insulators. Figure 4 shows Band structure for (a) $Bi_2Se_3$ (b) $Sr_{0.1}Bi_2Se_3$ (c) $Sr_{0.1}Bi_2Se_3$ at 2 GPa . Similarly in density of states (DOS) calculation for Sr doped entities, the density of states at Fermi energy is mainly contributed by Bi 5d, and Sr 4p orbital. This is also shown in figure 4 (d) $Bi_2Se_3$, (e) $Sr_{0.1}Bi_2Se_3$, and (f) $Sr_{0.1}Bi_2Se_3$ at 2 GPa. Total density of state about Fermi energy is around 3eV/state (per formula unit). For Sr intercalated $Bi_2Se_3$ total DOS($N_{Ef}$) at Fermi surface is 2.67 states/eV and at 2 GPa total DOS($N_{Ef}$) at Fermi Surface is estimated to be 2.43 states/eV. Evidently, the band-structure calculations also support the implication that decreasing $N(E_F)$ leads to decreasing $T_c$ under increasing external pressure.

Next we examine the normal state resistivity of $Sr_{0.1}Bi_2Se_3$ under hydrostatic pressure and its implication for electronic correlation. Towards this, we have fitted the low temperature resistivity data in the temperature range 8 to 28 K with the Fermi liquid dependence of resistivity; $\rho = \rho_0 + AT^2$. The excellent fitting of resistivity (inset (b) of figure 5)) shows applicability of Fermi liquid behaviour in $Sr_{0.1}Bi_2Se_3$ with hydrostatic pressure. The value of $\rho_0$ and A coefficient obtained from fitting are shown in the main panel and inset (a) of figure 5 respectively. Both $\rho_0$ and A increase with pressure. The coefficient of A which is a measure of electronic correlation, is directly proportional to square of the electronic effective mass $m^*$ [28].

Since A is increasing with pressure, $m^*$ is expected to increase. Further, since the density of states $N(E_F)$ goes as $\sim m^* n^{1/3}$ [18] and $m^*$ is increasing with pressure, the only way a net decrease in $N(E_F)$ can be conjured is by pressure induced decrease of carrier concentration $n$ at the Fermi level due to broadening of bands. Such behaviour effectively explains $T_c$ suppression in several low $T_c$ conventional elemental superconductors. In summary, both from high pressure measurements and from VASP calculations we find evidence for decreasing $N(E_F)$ leading to suppression of $T_c$ along with increased normal state resistivity.

Further, to understand the behaviour of resistivity in the temperature region above 100 K, under various pressures upto 8 GPa, $\rho(T)$ can be fitted with the small polaronic hopping model, $\rho = \rho_0 T \exp(E_a/T)$, [29] where $\rho_0$ and $E_a$ are the residual resistivity and activation energy for polaron hopping conduction respectively. We note that magnitude of resistivity is reasonably high and the addition of Sr onto the parent topological insulator $Bi_2Se_3$ is in some sense addition of disorder. We therefore need to consider behavior of Cooper pairs in a non-crystalline environment. From the slopes of $\ln(\rho/T)$ vs $(1/T)$ plot, $E_a$ was calculated for various pressures and it was observed that $E_a$ increased with application of pressure. Moreover, increase in $\rho(T)$ with temperature is consistent with scattering of degenerate electrons with phonons that are populated according to classical equi-partition distribution. The increase of resistivity can be understood by analysing activation energy. Further, the variable range hopping model also fits very well in the high temperature range with $\rho = \rho_0 \exp(T_0/T)^{0.25}$, [30] where $T_0$ is the constant. However, it should be noted that the temperature range where the VRH behavior appears to hold can have a significant overlap with the activated temperature range and also it is interesting to note that the high temperature polaronic state is increased by application of external pressure. A close analysis of the temperature dependence of the resistivity reveals the following; in a temperature range from 300 K down to 100 K, the electric conductance can be understood as a parallel circuit of an insulating bulk and a metallic surface. When the temperature is below 100 K, the resistivity tends to saturate and to deviate dramatically from the insulating behavior as described in the 3D VRH model. This indicates that a metallic transport channel plays a dominant role at low temperatures. Hence, the distinction between the two transport mechanisms can be a subtle issue that depends on the application of pressure.

## 4. Conclusion

In summary, the magnetic and electrical resistivity measurements under pressure are studied on the new $Sr_{0.1}Bi_2Se_3$ topological superconductor. The $T_c$ of $Sr_{0.1}Bi_2Se_3$ decreases monotonically with increasing pressure with the rate of -1.14 K/GPa as estimated from magnetization measurements. Band structure analysis involving external pressure to $Sr_{0.1}Bi_2Se_3$ shows decrease in DOS at Fermi level with application of pressure. The suppression of $T_c$ with increasing normal state resistivity and increasing electronic correlation is well accounted by decreasing $N(E_F)$ as evidenced in conventional low $T_c$ superconductors. Using polaronic and variable range hopping models, the residual resistivity and activation energy are shown to increase with application external pressure in the normal state that is reminiscent of electronic transport in the presence of intercalated disorder.

## 5. Acknowledgements

KM thanks UGC-RGNF, University Grant Commission, India and GK thanks UGC-SAP,

University Grant Commission, India for fellowships. SA acknowledges DST (SERB, FIST, PURSE), BRNS, CEFIPRA, DRDO and UGC (SAP), New Delhi. Shruti thanks University Grant Commission, India for her SRF fellowship. SP thank FIST program of Department of Science and Technology, Government of India for low temperature high magnetic field facilities at JNU. We thank Dr. Ambesh Dixit (IIT Jodhpur) for useful discussions.

**Figure Captions:**

**Figure 1:** Resistivity as function of temperature at different pressure values from 0 to 8 GPa. The inset shows the resistivity in low temperature range showing decrease in $T_c$ with increasing pressure.

**Figure 2:** Temperature dependent magnetization (M(T)) subtracted with magnetization at 3 K (M(3K)) in ZFC mode of $Sr_{0.1}Bi_2Se_3$ at different pressure from 0 to 0.99 GPa. Inset shows the magnetization indicating $T_c$ of $Sr_{0.1}Bi_2Se_3$ below 3 K (marked with arrow). The superconducting transition of Sn manometer is shown inside ellipsoidal curve above 3 K.

**Figure 3:** Variation in superconducting transition temperature $T_c$ with pressure obtained from magnetization measurement. The red line shows the extrapolation using polynomial fit to the $T_c(p)$ determined from magnetization data. The inset shows the variation of $\rho_{3.5K}$ and $\rho_{300K}$ with pressure.

**Figure 4:** Electronic band structure of (a) $Bi_2Se_3$ (b) $Sr_{0.1}Bi_2Se_3$ (c) $Sr_{0.1}Bi_2Se_3$ at 2 GPa. The corresponding Density of States are plotted in (d), (e), and (f) respectively.

**Figure 5:** Main panel shows residual resistivity $\rho_0$ as a function of pressure. Inset (a) shows the variation the coefficient A as a function of pressure. Inset (b) shows an excellent fitting to the resistivity from 8 to 28 K using $\rho = \rho_0 + AT^2$.

**Figure 6:** Fitting of (a) small polaron model [ln ($\rho/T$) vs 1/T] and (b) VRH model [ln $\rho$ vs $1/T^{0.25}$] for the high temperature resistivity above 100 K.

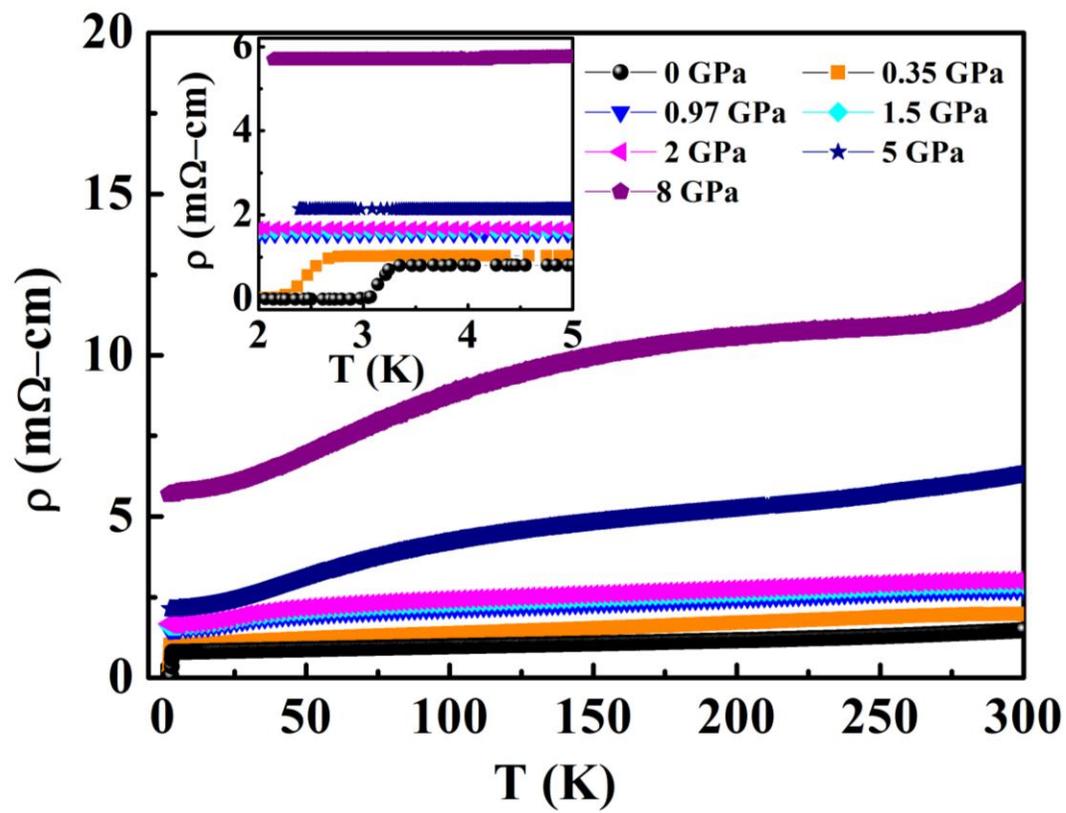

**Figure 1**

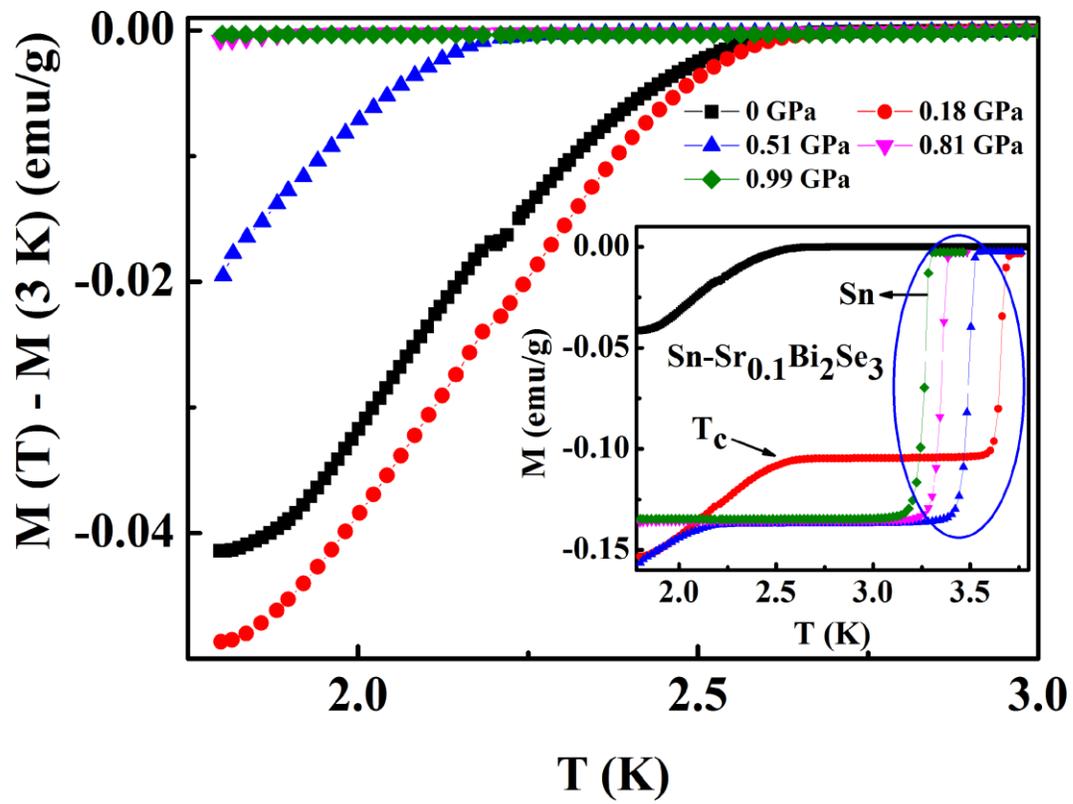

**Figure 2**

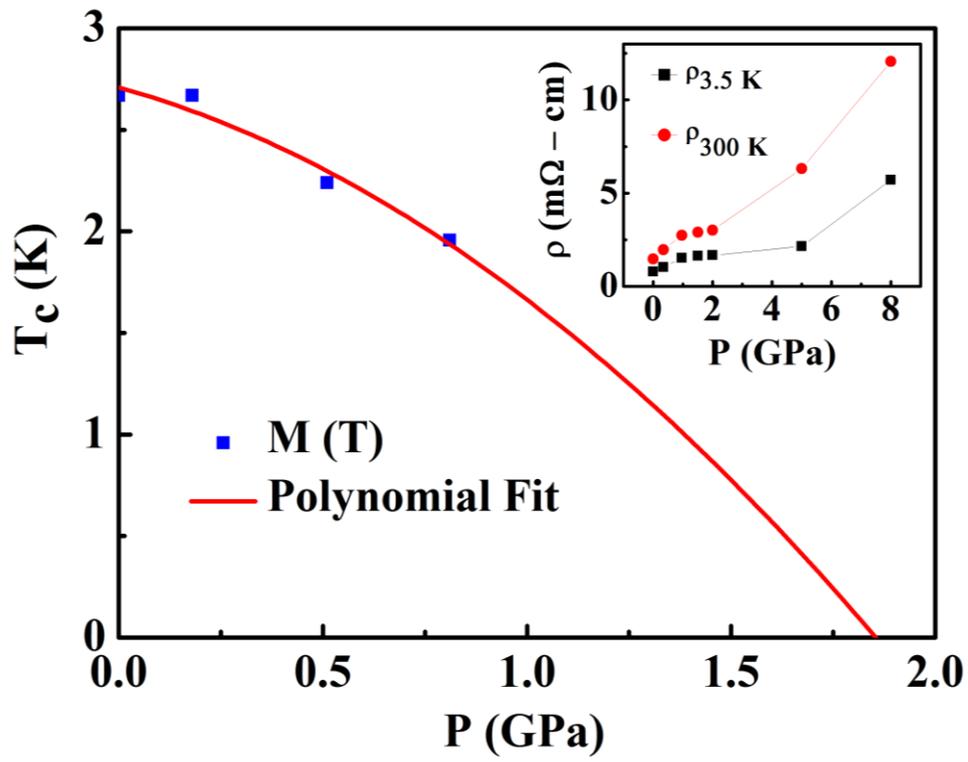

**Figure 3**

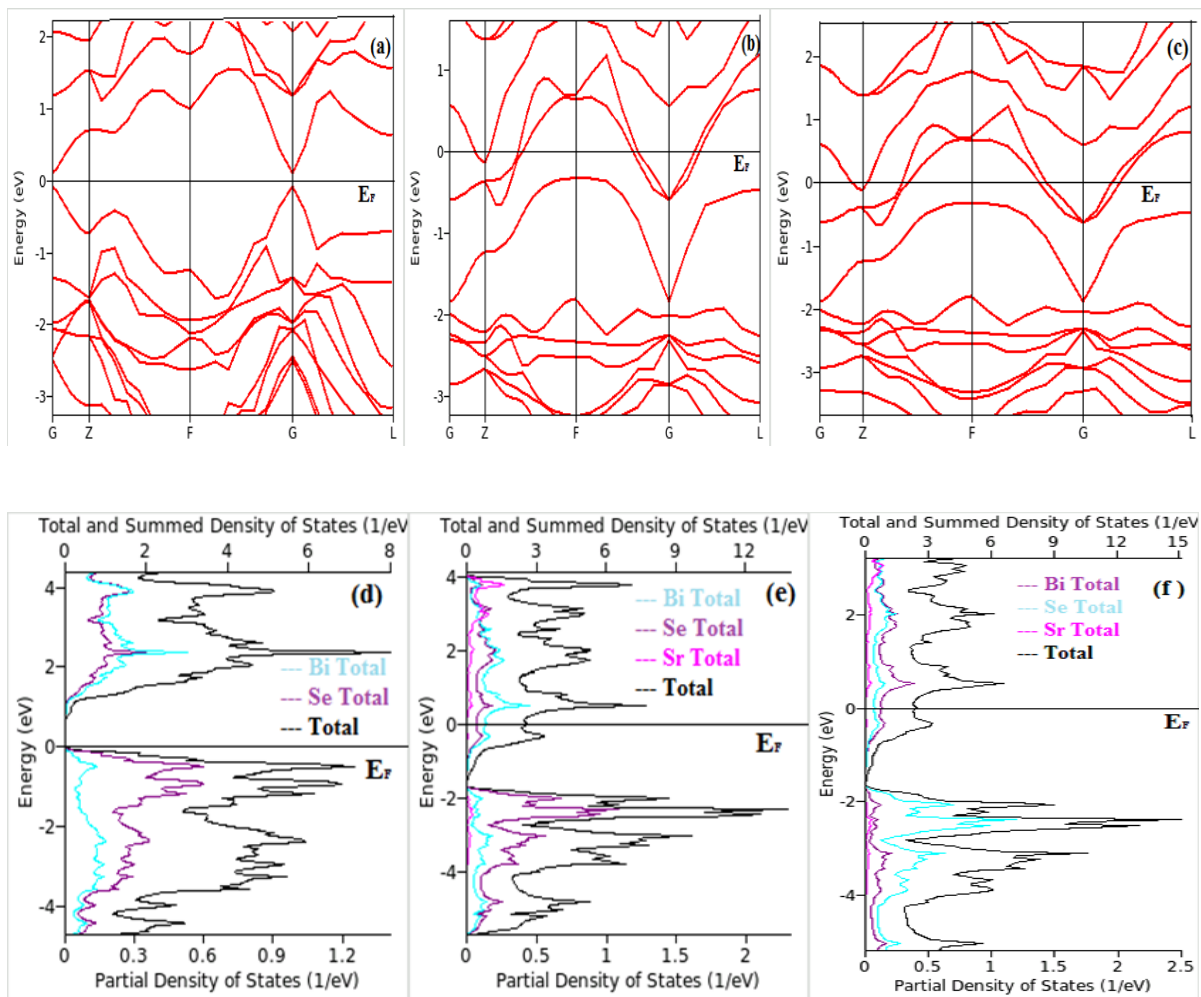

**Figure 4**

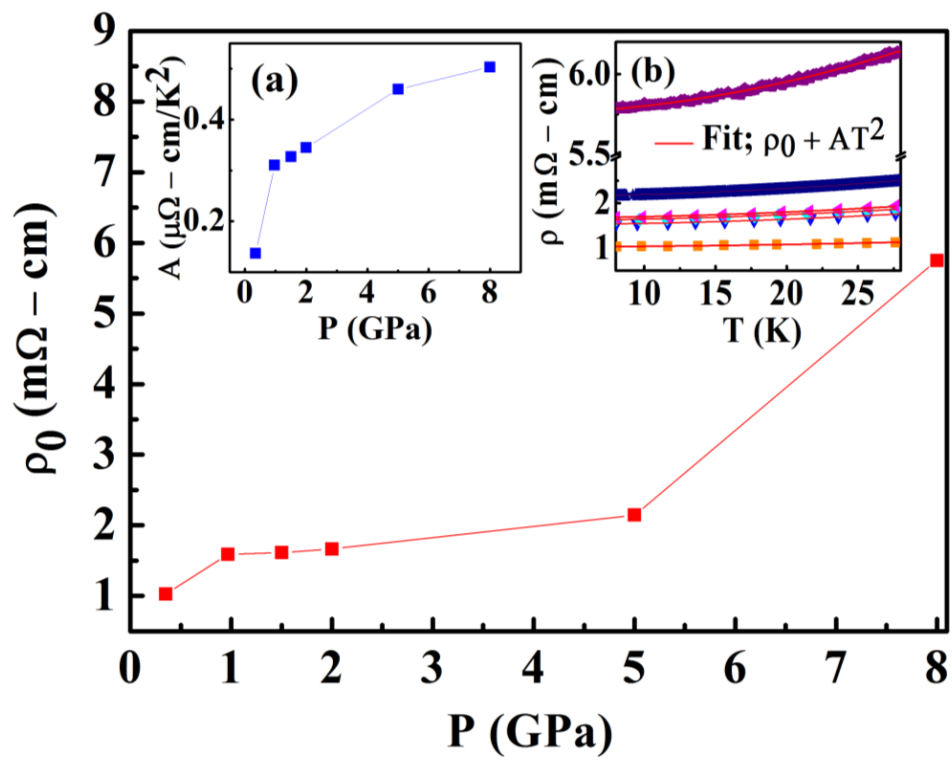

Figure 5

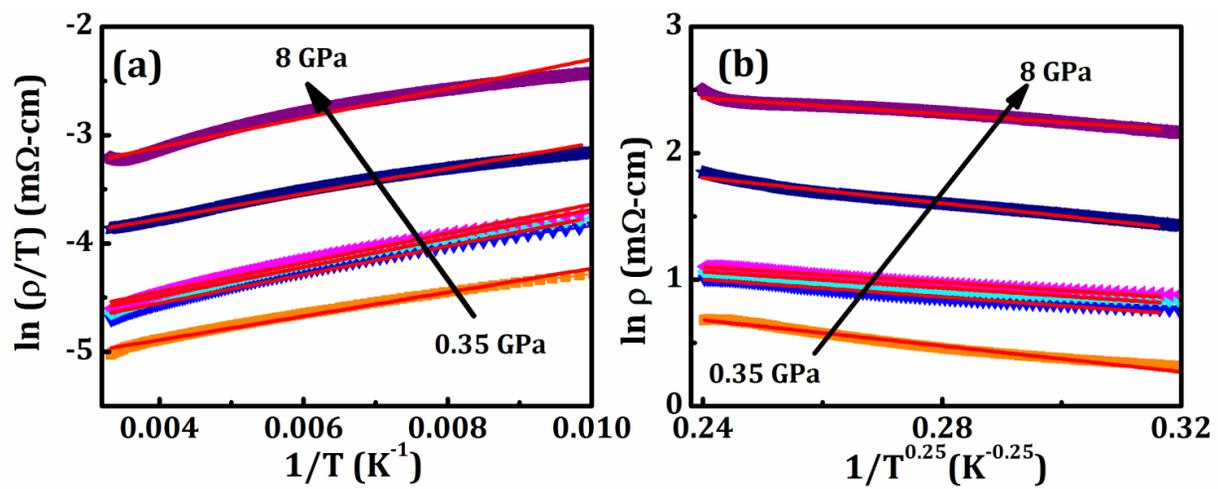

**Figure 6**